# Coherence of a field-gradient-driven singlet-triplet qubit coupled to many-electron spin states in $^{28}$Si/SiGe

Younguk Song[1†], Jonginn Yun[1†], Jehyun Kim[1], Wonjin Jang[1], Hyeongyu Jang[1], Jaemin Park[1], Min-Kyun Cho[1], Hanseo Sohn[1], Noritaka Usami[2], Satoru Miyamoto[2], Kohei M. Itoh[3], and Dohun Kim[1]*

[1]Department of Physics and Astronomy, and Institute of Applied Physics, Seoul National University, Seoul 08826, Korea

[2] Graduate School of Engineering, Nagoya University, Nagoya, Japan

[3] Department of Applied Physics and Physico-Informatics, Keio University, Yokohama, Japan

[†]These authors contributed equally to this work.

*Corresponding author: dohunkim@snu.ac.kr

Engineered spin-electric coupling enables spin qubits in semiconductor nanostructures to be manipulated efficiently and addressed individually. While synthetic spin-orbit coupling using a micromagnet is widely used for driving qubits based on single spins in silicon, corresponding demonstration for encoded spin qubits is so far limited to natural silicon. Here, we demonstrate fast singlet-triplet qubit oscillation (~100 MHz) in a gate-defined double quantum dot in $^{28}$Si/SiGe with an on-chip micromagnet with which we show the oscillation quality factor of an encoded spin qubit exceeding 580. The coherence time $T_2$* is analyzed as a function of potential detuning and an external magnetic field. In weak magnetic fields, the coherence is limited by fast noise compared to the data acquisition time, which limits $T_2$* < 1 µs in the ergodic limit. We present evidence of sizable and coherent coupling of the qubit with the spin states of a nearby quantum dot, demonstrating that appropriate spin-electric coupling may enable a charge-based two-qubit gate in a (1,1) charge configuration.

Balancing the manipulation speed and coherence time, which often play opposing roles, has been a major goal of semiconductor quantum dot-based quantum information processing platforms[1–3] to maximize the qubit control fidelity. The electrical control of spin states is a representative example where, depending on the properties of the host material, either intrinsic[4,5] or extrinsic[6,7] spin-electric coupling methods have been explored. While strong spin-orbit coupling in compound semiconductors such as InAs and InSb enables fast Rabi oscillations[4,5], uncontrolled and substantial susceptibility to charge noise limits the inhomogeneous coherence time $T_2^*$ to the order of tens of nanoseconds. More recently, hole spins in group IV materials such as Ge [8] and Si [9] or electron spins in the Si-MOS structure[10] have been attracting much attention due to a more favorable ratio between the spin-orbit-based control speed and coherence time.

The electrons in silicon, in particular in the Si/SiGe heterostructure, have small intrinsic spin-orbit coupling[11]; therefore, an extrinsic method such as a micromagnet is necessary to rapidly manipulate its spin states. For single-spin qubits, the placement of an on-chip micromagnet has proven effective for both natural[7,12] and isotopically enriched silicon[13] in Si/SiGe and Si-MOS structures, where the field gradient provides fast control while not severely compromising the spin coherence. In the case of the silicon-based two-electron singlet-triplet qubit, however, the efficiency of the technique involving a micromagnet has not been fully examined. Previous studies of singlet-triplet qubit operation either used a small field gradient[14] or relied on the modulation of the exchange energy[15] in natural silicon. Exploration of the micromagnet technique with a field gradient in the intermediate range in isotopically purified silicon would thus be important for optimizing spin-electric coupling. In addition, this approach would enable this route to be compared with other methods such as the recently demonstrated spin-valley-driven coherent singlet-triplet oscillation in silicon[16,17].

Here, we demonstrate singlet-triplet qubit oscillation in a gate-defined double quantum dot in $^{28}$Si/SiGe. An on-chip micromagnet is used to generate a magnetic field gradient that is sufficient to allow fast manipulation (oscillation frequency $f_Q \sim$ 100 MHz), while benefiting from high spin coherence by isotopic enrichment. We measure the variation in the spin-electric coupling strength in the large valley-splitting regime (> 175 μeV) in which an appropriate field gradient enables an encoded spin qubit to attain an oscillation quality factor over 580. We also present the analysis of the variation in $T_2^*$ as a function of experimental parameters such as detuning $\varepsilon$, magnetic field $B_{z,\text{ext}}$, and gate tuning conditions, exploring the origin of the dominant noise source in the system. Moreover, we present evidence that the qubit engages in sizable and coherent coupling with the spin states of a nearby quantum dot, thereby demonstrating that the appropriate amount of spin-electric coupling may enable a novel type of two-qubit gates of encoded spin qubits.

**The triple quantum dot system**

Figure 1a shows a multiple quantum dot device fabricated on top of a $^{28}$Si/SiGe heterostructure (see Methods for details of the material structure and device fabrication). We focus on a two-electron singlet-triplet (ST$_0$) qubit formed by the gate electrodes near the left Ohmic contact and a global top gate (not shown) while the regions beneath the other electrodes are fully accumulated. The general Hamiltonian $H$ of the ST$_0$ qubit can be expressed as $H = J(\epsilon)\sigma_z + \Delta B_z \sigma_x$, where $J(\varepsilon)$ is a $\varepsilon$-dependent exchange interaction with the Pauli matrix $\sigma_{i=x,y,z}$. $\Delta B_z$ is the magnetic field difference between the quantum dots constituting the qubit and is denoted in the frequency unit Hz using Tg$\mu_B$/h, where g, $\mu_B$, and h are the Lande g-factor of the electrons in silicon, the Bohr magneton, and Planck's constant, respectively.

Additionally, we formed a third, many-electron quantum dot next to the ST$_0$ qubit to study the capacitive interaction between them. The design of the micromagnet on top of the is similar to the ones used previously[18]. High frequency and synchronous voltage pulses, combined with the DC voltage through bias tees, were input to gates $V_1$, $V_2$, and $V_T$. Fast RF reflectometry[19,20] was performed by injecting a carrier signal with a frequency of approximately 125 MHz and power of –100 dBm at the Ohmic contact of the RF single-electron transistors on the left (see Fig. 1a). The reflected power was monitored through a chain of cryogenic and room temperature amplification and subsequent homodyne detection. The device was operated in a dilution refrigerator with a base temperature of approximately $\approx 7$ mK, with $B_{z,\text{ext}}$ ranging from –400 mT to 400 mT applied in the direction shown in Fig. 1a.

Figure 1b shows the charge stability diagram of the ST$_0$ qubit coupled with a many-electron quantum dot. The full specification of the number of electrons in the left, middle, and right quantum dots (QD$_L$, QD$_M$, QD$_R$, see green circles in Fig. 1a) are denoted as (n,m,l), whereas the (n,m) notation is used whenever we focus on the ST$_0$ qubit only. A voltage pulse with a width of approximately 10 ns and rise time of 0.5 ns is input to $V_1$ and $V_2$ in the directions indicated by the red arrows in Fig. 1b. Near the charge transition from (0,2,N) to (1,1,N), the pulse abruptly changes the Hamiltonian to the form $H = \Delta B_z \sigma_x$, where the spin state initialized to the singlet rotates around the *x*-axis on the Bloch sphere (Larmor oscillations[11]), thereby resulting in a non-zero triplet state probability $P_T$. The discrimination of the resultant excited state population is conventionally performed by Pauli spin-blockade (PSB)-based spin-to-charge conversion[2,11,21] where the singlet and triplet spin states are mapped to the (0,2) and (1,1) charge configurations, respectively. However, $\Delta B_z$ produced by the micromagnet facilitates relaxation of the transient triplet (1,1) to the singlet (0,2) by mixing with the singlet (1,1) state, which makes a high-fidelity single-shot readout problematic[22].

To circumvent the problem, we adopted one of the latched-PSB techniques that maps the triplet state to a long-lived metastable charge configuration[23–25]. Pioneered in a similar experiment performed in a GaAs triplet quantum dot system[24], the version we used converts the triplet state (1,1) to the (1,2) state by rapidly loading an electron from the reservoir at a tunneling rate greater than the sensor bandwidth of 10 MHz (Fig. 1c, middle panel, see also inset in Fig. 1b). On the other hand, tunneling to the reservoir on the left is tuned to be of the order of 10 Hz. At this rate, the metastable (1,2) state can relax to a singlet (0,2) state only by indirect and slow tunneling of an electron from $QD_L$ to the reservoir on the right (Fig. 1c, rightmost panel). Along with the higher signal contrast of a charge of one electron compared with conventional PSB, the prolonged relaxation time of the triplet states enables fast and high-fidelity single-shot measurement. The fast measurement capability is also important to examine the extent to which the variation in the qubit coherence time depends on the total data acquisition time to determine the effect of slow charge noise[26–28], as discussed in detail below.

Figure 1d shows the magnetospectroscopy measurements of the valley splitting[29] for $QD_L$ and $QD_M$. By observing the crossover of the ground state from the singlet to the triplet by measuring the dependence of the energy required to add the second electron to each dot on $B_{z,ext}$, we obtain the valley splitting ~ 175 μeV (257 μeV) in $QD_L$ ($QD_M$). The result confirms that the valley splitting in our device is at the largest energy scale of at least twice that of the Zeeman splitting at the maximum $B_{z,ext}$ applied in this study. Thus, we ignore the valley degree of freedom in this work and focus only on the $\Delta B_z$-driven $ST_0$ qubit dynamics.

**Qubit dynamics driven by the field gradient**

With the calibrated π/2 pulse obtained from the Larmor oscillation measurement at the pulse amplitude $|V_1| = |V_2| = 270$ mV, we construct a three-step pulse sequence for Ramsey interferometry (Fig. 2a). During the second step, at the pulse amplitude of free evolution $V_{evol}$, the qubit evolves around the axis of the Bloch sphere determined by the ratio of $J(V_{evol})$ and $\Delta B_z$. Figure 2b shows a representative quantum oscillation at $V_{evol} = 770$ mV under the representative tuning conditions. This demonstrates a record-high[11,14,15] oscillation quality factor $Q^* = f_Q \times T_2^* = 116.25$ MHz × 4.8 μs = 558 of a $\Delta B_z$-driven $ST_0$ qubit rotation in the deep (1,1) charge configuration. In addition, high-resolution measurement (10,000 shots per data point with a single-shot readout time of 20 μs) of the first few oscillations (Fig. 2b, left panel) shows a readout visibility of ~ 85% (see Supplementary Information S1 for details on the signal-to-noise ratio).

To more fully understand spin-electric coupling and its effect on the coherence time, we mapped the dependence on the free evolution time $t_{evol}$ and $V_{evol}$ of the Ramsey interference at $B_{z,ext} = 300$ mT, as shown in Fig. 2c. The oscillation observed for $V_{evol} < 0.1$ V shows the fast but short-lived oscillation driven by $J$ (marked as ● in Fig. 2d and the corresponding schematic diagram in Fig. 2e), whereas the oscillations driven by $\Delta B_z$ exhibit prolonged $T_2^*$ (white dashed contour in Fig. 2c) for $V_{evol} > 0.1$ V due to the lower charge noise susceptibility $df_Q/dV_{evol}$. In this regime, the data in the time and frequency domains exhibit the following main features. First, $f_Q$ is generally linearly dependent on $V_{evol}$, which arises from the presence of the micromagnet (see ■ and ★ in Fig. 2d), and this is consistent with the previously observed linear shift of the single-spin resonance frequency in silicon in the presence of the synthetic field gradient[13]. Second, $T_2^*$ depends non-monotonically on $V_{evol}$. In particular, a significant decrease in $T_2^*$ is observed in the vicinity of $V_{evol} = 0.35$ V (near ⬥ in Fig. 2d). Third, $f_Q$ undergoes an abrupt frequency shift of about $\Delta f_Q \sim 1.7$ MHz at approximately $V_{evol} = 0.45$ V

(▲ in Fig. 2d). Estimated from the calibrated lever arm of 0.023, the cross-talk effect of $V_{evol}$ = 0.45 V on $QD_R$ shifts the chemical potential of $QD_R$ to the Fermi-level of the right contact $E_F$ where the ground state charge transition occurs. Therefore, the observed $\Delta f_Q \sim 1.7$ MHz per one electron change is the measurement of the capacitive coupling between the $ST_0$ qubit and $QD_R$. We additionally verified this interpretation by adjusting the DC tuning of the plunger gate of $QD_R$ and by observing the systematic shifts of the point ▲ (see Supplementary Information S2).

In general, the charge fluctuation in $QD_R$ adversely affects the coherence of the capacitively coupled qubit. However, we note that $V_{evol}$ = 0.35 V (near ⬥ in Fig. 2d) showing the lowest $T_2^*$ occurs below $V_{evol}$ = 0.45 V (▲ in Fig. 2d) where $QD_R$ experiences the maximum charge fluctuation. Assuming that non-negligible spin-dependent coupling occurs between the $ST_0$ qubit and the Zeeman-split ground and excited spin states in $QD_R$ occupied by $N$ electrons ($E_g$ and $E_e$, respectively), the qualitative interpretation of this phenomenon is as follows (See Supplementary Information S3 for data supporting spin-dependent coupling). For detuning near ■, the spin state in $QD_R$ remains in the ground state with high fidelity as both $E_g$ and $E_e$ are well below $E_F$, and the spin-dependent charge fluctuation is low. From point ■ to ▲, as $E_e$ approaches and passes $E_F$, fast tunneling between $QD_R$ and the reservoir through $E_e$ leads to enhanced spin-dependent charge number fluctuation. This fluctuation reduces $T_2^*$ and produces a small frequency shift of $\Delta f_Q \sim 0.4$ MHz (red arrows in Fig. 2d). In this regime, the spin state in $QD_R$ is expected to be a mixed state because of the non-negligible average occupation in $E_e$. The maximum charge fluctuation, hence the minimum $T_2^*$, is expected to occur when $E_e$ approximates $E_F$, which corresponds to the point ⬥. Finally, at point ▲, $E_g$ aligns with $E_F$, resulting in the change in the full electron number in $QD_R$ and the appearance of the kink in time-averaged $f_Q$ measurement.

More quantitatively, we compared the experimental results and those of the numerical simulation[30] using the following phenomenological Hamiltonian and Lindblad operators.

$$H = J(V_{evol})\sigma_z \otimes \mathbf{1} + \Delta B_z(V_{evol})\sigma_x \otimes \mathbf{1} + \beta(V_{evol})\mathbf{1} \otimes \sigma_z + \gamma(B_{z,ext})\mathbf{1} \otimes \sigma_x$$
$$+ J_{int}(V_{evol}) \frac{1}{e^{\eta(\beta(V_{evol}-c))}+1} \sigma_x \otimes \sigma_z$$
$$L_1 = \tau_1 \sqrt{J(V_{evol})} \sigma_z \otimes \mathbf{1}, L_2 = \tau_2 \sqrt{\beta(V_{evol})} \mathbf{1} \otimes \sigma_z$$

Here, the Hamiltonian describes the two interacting qubits, $ST_0$ qubit, and the two spin states of the nearby $N$-electron quantum dot, $QD_R$. More specifically, the Hamiltonian of the $ST_0$ qubit is constructed as $J(V_{evol})\sigma_z + \Delta B_z(V_{evol})\sigma_x$, where the background $\Delta B_z(V_{evol}) = 47.7\ V_{evol} + 61.3$ ( $0.25\ V < V_{evol} < 0.7\ V$ ) is estimated from Fig. 2d. We assumed that the Hamiltonian of $QD_R$ is analogous to that of the $ST_0$ qubit, such that the diagonal term $\beta(V_{evol})$ is an exponential function of $V_{evol}$, and the off-diagonal term $\gamma(B_{z,ext})$ is a linear function of $B_{z,ext}$. Furthermore, the coupling we consider is the spin-electric coupling induced from the spatial distribution of the orbital wavefunction of $QD_R$ depending on its spin states. In view thereof, we chose the spin-electric coupled eigenstate of $QD_R$ as the $\sigma_z$ basis and introduced the interaction between the $ST_0$ qubit and $QD_R$, which was assumed to be in the form of $\sigma_x \otimes \sigma_z$. This coupling term is multiplied by a phenomenological Fermi–Dirac distribution with proper constant c, $\eta$ and coupling strength $J_{int}(V_{evol})$ and to incorporate the change in the charge state of the nearby $N$-electron quantum dot (see Supplementary Information S4 for details of the simulation). Additionally, the phenomenological Lindblad operators for the $ST_0$ qubit ($L_1$) and two-level system in $QD_R$ ($L_2$) are introduced with the proportionality constant $\tau_1$ ($\tau_2$) to reflect the experimentally observed decoherence.

The inset in Fig. 2c shows the simulation result which consistently reproduces the sudden kink in the frequency near $V_{evol} = 0.45$ V, the significant decrease in $T_2^*$ near $V_{evol} =$

0.35 V, and the subsequent recovery of $T_2^*$ near the kink. Overall, by comparing the result of the simulation with that of the experiment, we concluded that the kink in the frequency and the drop in $T_2^*$ indicate capacitive coupling between the $ST_0$ qubit and $QD_R$ during the charge transition of $QD_R$.

**Coherent coupling between ST₀ qubit and many-electron spin states**

We further substantiated the validity of the above analysis by showing that the experimental and simulation results were consistently comparable under different quantum dot tuning conditions. Specifically, we fine-tuned the gate voltage levels to induce significant deviations in the overall coupling strength and the decoherence rates compared with the previous tuning condition. In this new tuning, we observed the characteristic beating of the quantum oscillation below $V_{evol} < 0.42$ V, as shown in Fig. 3a. Notably, the coherence of the oscillation markedly diminished when the coupling between the two qubits became appreciable. Figure 3b enables a more detailed examination of these results and provides the line cuts that offer a clearer comparison between the oscillation traces in the uncoupled ($V_{evol} = 0.6$ V, top trace in Fig. 3b) and coupled ($V_{evol} = 0.2$ V, bottom trace in Fig. 3b) regimes.

The manifestation of the beating oscillation under this specific tuning condition suggests that two different qubit frequencies emerge depending on the spin state in $QD_R$, and that the interqubit coupling rate is faster than the decoherence rate, unlike the previous tuning. In addition, the beating of the oscillation suggests that, below certain $V_{evol}$ values, the eigenstate of $QD_R$ becomes the superposition of the states of which the orbital wavefunction directly couples with $f_Q$ through the spin-electric coupling. The significant drop in $T_2^*$ in the regime of sizeable coupling is again likely a consequence of the interplay between the interqubit coupling and the dephasing effect discussed in the previous section. Note that the coupling strength with the maximum value of approximately 10 MHz also exhibits a dependence on $V_{evol}$, thereby highlighting the electrical tunability of the interqubit coupling strength.

This experimental result was compared with the numerical simulation, which employed the identical Hamiltonian and Lindblad operators introduced in the previous section, whose parameters were appropriately adjusted to reflect the different tuning conditions. One of the key adjustments involves the parameters of $\beta(V_{evol})$, which effectively transform the eigenstate of $QD_R$ into the superposition of the $\sigma_z$-eigenstates for $V_{evol} < 0.4$ V, which gives rise to the observed beating of the oscillation through the spin-electric coupling $\sigma_x \otimes \sigma_z$ term. The numerical calculation consistently reproduces the experimental results, including the characteristic kink near $V_{evol} = 0.48$ V, the emergence of the beating, and the significant reduction in $T_2^*$ throughout the coupling regime. These observations can be attributed to the mixed eigenstate of $QD_R$ and the Lindblad operators, which effectively mimic the aforementioned dephasing effect. Overall, our spin-electric coupling scenario convincingly reproduces the experimental results for various coupling parameters.

**External field dependence of $ST_0$ qubit coherence**

We turn to discuss the dominant noise source limiting the coherence of $ST_0$ qubit by investigating the variation in $f_Q$ and $df_Q/dV_{evol}$ as a function of $B_{z,ext}$. The magnitude of the field gradient $|\Delta B_z|$ is determined by measuring $f_Q$ of the $\Delta B_z$-dominated Ramsey oscillations at $V_{evol}$ = 800 mV for various values of $B_{z,ext}$ from 400 mT to –400 mT (Fig. 4a). Generally, $|\Delta B_z|$ was positively correlated with $B_{z,ext}$, which likely originated from the formation of multiple domains due to the demagnetization of the Co micromagnet at low $B_{z,ext}$. The calculated value expected for $|\Delta B_z|$ by simulation of the magnetic field using the Object Oriented Micromagnetic

Framework (OOMMF)[31,32] was in qualitative agreement with the experimental observation. (see Supplementary Information S5 for details of the micromagnetic simulation).

The controllability of $|\Delta B_z|$ via $B_{z,ext}$ paved the way to test whether a decrease in $|\Delta B_z|$ could lead to a smaller $df_Q/dV_{evol}$ and, consequently, an improved $T_2^*$ at low $B_{z,ext}$. Figure 4b shows the dependence of $df_Q/dV_{evol}$ on $f_Q$, extracted at various levels of $B_{z,ext}$ and $V_{evol}$. Unexpectedly, a strong correlation did not exist between $f_Q$ and $df_Q/dV_{evol}$. Depending on the experimental iteration, $df_Q/dV_{evol}$ was widely dispersed even at similar $f_Q$ controlled by $V_{evol}$. We again attribute this to the nanoscale formation of multiple domains in the micromagnet, which generates a locally inhomogeneous field distribution.

Figure 4c shows $T_2^*$ and $Q^*$ as functions of $B_{z,ext}$ at several $V_{evol}$. Generally, the decreasing $Q^*$ is predominantly the result of the rapid decrease in $f_Q$ as the applied magnetic field $B_{z,ext}$ weakens, whereas $T_2^*$ varies at most by a factor of two as a function of $B_{z,ext}$. The latter finding is also consistent with the observation that $f_Q$ and $df_Q/dV_{evol}$ are not strongly correlated. Although $T_2^*$ tends to increase in the presence of strong $B_{z,ext}$, we argue that this is because of the interplay between the experimental data acquisition time and dominant noise band, which shifts to the low frequency at stronger $B_{z,ext}$. We confirm that extending the total data acquisition time significantly affects $T_2^*$ at $B_{z,ext}$ = 400 mT, indicating that slow charge noise compared to a given measurement time plays an important role (see Supplementary Information S6). Moreover, near $V_{evol}$ = 350 mV, where $T_2^*$ is limited by strong coupling with the spin states in $QD_R$ and fast charge noise is therefore presumed to dominate the noise spectrum, $T_2^*$ is nearly constant as a function of $B_{z,ext}$. In this respect, we assume that $T_2^*$ at low $B_{z,ext}$, which approximates 1 μs regardless of the value of $V_{evol}$, is entirely dominated by the noise spectrum, which is faster than the measurement time. This indicates that the coherences of the $ST_0$ qubit are closer to the ergodic limit.

The dominance of high-frequency noise in our system is also supported by measurement of the spin-echo time $T_{echo}$ and echoed quality factor $Q_{echo}$ at various $B_{z,ext}$ and $V_{evol}$ (Fig. 4d). Notably, the spin-echo enables only a minor improvement in the coherence time by a factor of at most two compared with $T_2^*$ at low $B_{z,ext}$ of ~100 mT, thereby indicating that the major source of noise in this regime is in the high-frequency band. Similar ineffectiveness of the spin-echo was observed in a $^{Nat}Si/SiGe$-based singlet-triplet qubit for non-negligible $J$ [33]. Although a previous study[34] pointed out that the increased flip-flop motion of residual $^{29}Si$ nuclear spins at low $B_{z,ext}$ leads to the reduction of $T_2^*$, we rule out this possibility since this quasi-static noise would have been effectively corrected by the echo sequence. Moreover, this type of noise is more likely to occur under RF excitations needed for single-spin qubit manipulation. The absence of such control in this experiment also indicates that the mechanism of the dominant noise source at low $B_{z,ext}$ in our experiment differs from that in the previous study. The spin-echo more effectively enhances the coherence time at high $B_{z,ext} > 300$ mT, which is consistent with our scenario that, in this regime, the dominant noise primarily stems from the low-frequency band. Similar to the behavior of $T_2^*$, the spin-echo is not effective when the $ST_0$ qubit is strongly coupled with $QD_R$ (see Fig. 4d, third panel).

We additionally compared the power spectral density (PSD) of noise for strong and weak magnetic fields, $B_{z,ext}$ (400 and 50 mT, respectively) obtained by the single-shot measurement-based rapid Bayesian estimation method[26,35,36] (see also Supplementary Information S7) as shown in Fig. 5a. Although both of these spectra exhibit a larger white noise component compared to previous studies[28,33], the PSD of the spectrum at $B_{z,ext} = 50$ mT is about two orders of magnitude larger across the entire range of frequencies with different exponent α of the $1/f^α$-like power spectrum compared to the PSD at $B_{z,ext} = 400$ mT. Assuming that the frequency-independent noise extends to frequencies beyond the experimentally measured limit of ~40 Hz, the result explains the overall ineffective noise refocusing via the spin-echo technique in our system, in particular at low $B_{z,ext}$. In addition, the tendency of the white noise floor to increase with decreasing $B_{z,ext}$ along with the change in α (Fig. 5b), which is generally indicative of a relative increase in the portion of fast charge noise, is consistent with the variation in $T_2^*$ and $T_{echo}$, as presented in Fig. 4.

**Discussion and conclusions**

The origin of the rather high white noise floor in our system, which further increases at low $B_{z,ext}$, remains an open question. Although a more comprehensive understanding of the dominant noise source would require further experiments, the ineffective coherence recovery using the spin-echo technique due to relatively fast noise indicates that the noise does not predominantly originate from the increased flip-flop rate of the residual $^{29}$Si nuclear spins. Based on our investigation of the signature of the nanoscale multi-domain structure as the micromagnet demagnetizes at $B_{z,ext}$ < 200 mT, we speculate that the fast noise could have stemmed from the interplay between the field inhomogeneity induced by the magnetic domain structure and charge noise. This could be clarified by studying multiple devices containing micromagnets with various magnetic properties. A potential approach could involve the use of different techniques for micromagnet fabrication; for example, the deposition of magnetic material in the presence of an applied magnetic field, which is known to induce a preferential magnetization axis and hence a significantly modified hysteresis loop[37]. This technique may enable the magnetic structure to be more stably controlled in a weak magnetic field, which would allow an investigation of the transduced noise with varying magnetic properties. Moreover, the technique could also be useful for other applications such as semiconductor-superconductor hybrid circuits[38,39] for long-range coupling where operation in a weak magnetic field is beneficial.

Nevertheless, we successfully demonstrated coherent $ST_0$ oscillations with outstanding $Q^*$. This was enabled by using an on-chip micromagnet technique in an isotopically purified $^{28}$Si/SiGe heterostructure where $f_Q$ is tunable in the (1,1) charge configuration due to the dependence of the magnetization on $B_{z,ext}$. Our findings reveal that capacitive coupling can facilitate coherent interactions between two quantum systems: the two-electron $ST_0$ qubit and the many-electron quantum dot. Moreover, by formulating Hamiltonians for these quantum systems and their interactions, we effectively reproduced the coherent $ST_0$ qubit oscillation observed in our experiments through numerical simulation. Our work also suggests areas for improvement. Even though our device was designed to allow us to coarsely tune the chemical potential of $QD_R$, independent control of the quantum states of $QD_R$ was challenging because of the limited number of control lines in the current single-gate layer structure. Enhanced control over individual quantum dots and precise coupling strength modulation could be attained by adopting an overlapped gate structure[40], which may enable novel two-qubit gate schemes for encoded spin qubits in silicon.

**Methods**

**Material structure and device fabrication**

The $^{28}$Si/SiGe heterostructure wafer was grown by a molecular beam epitaxy growth method. An isotopically purified silicon source (with a residual $^{29}$Si concentration of approximately 800 ppm) was used for the strained quantum well with a thickness of 12 nm. The design of the surface gate electrode resembles that of GaAs spin qubit devices where both quantum dot confinement and barrier gates reside in the same layer and a global accumulation gate is used for electrostatic doping. The dimensions of the accumulation gate were maintained below 2 x 2 μm$^2$ to minimize the parasitic capacitance[20], enabling proper impedance matching conditions for radio frequency (RF) reflectometry. A Co micromagnet was deposited above the accumulation gate using an e-beam evaporator, with a Au cap for the antioxidation layer.

**Measurement setup**

The sample was cooled to the base temperature, ~7 mK, with a cryogen-free dilution refrigerator (Oxford Instruments Triton-500). A sensing dot based on an RF single-electron transistor was used to detect the change in the charge state of $QD_L$, $QD_M$, and $QD_R$ in our system. An onboard inductor of 1500 nH and a parasitic capacitance on the order of 1 pF formed an LC-tank circuit with a resonance frequency at ~125 MHz, which was used for RF reflectometry. Two arbitrary waveform generators (HDAWG and Operator-X+ by Zurich Instruments and Quantum Machines, respectively) were used to synchronize the multi-channel voltage pulses and timing marker generation. A high-frequency lock-in amplifier (Zurich Instruments, UHFLI) was used as a carrier generator and demodulator for homodyne detection. At room temperature, a carrier power of –40 dBm was generated which was further attenuated by –50 dB by the cryogenic attenuators and the directional coupler. The reflected signal is initially amplified by 50 dB with the cryogenic amplifier (Caltech Microwave Research Group, CITLF2 x2 in series), and then additionally amplified by 20 dB at room temperature using a custom-built RF amplifier. We used the QUA (Quantum Machines) language framework for scripting experimental sequences, performing single-shot readouts, and signal conditioning.

**Data availability**

The data that support the findings of this study are available from the corresponding author upon request.

**Acknowledgments**

This work was supported by a National Research Foundation of Korea (NRF) grant funded by the Korean Government (MSIT) (No. 2019M3E4A1080144, No. 2019M3E4A1080145, No. 2019R1A5A1027055, RS-2023-00283291, SRC Center for Quantum Coherence in Condensed Matter RS-2023-00207732, and No. 2023R1A2C2005809) and the core center program grant funded by the Ministry of Education (No. 2021R1A6C101B418). The work on the $^{28}$Si/SiGe growth was supported by JST Moonshot R&D grant No. JPMJMS226B and JSPS Grant-in-Aid for Scientific Research (KAKENHI) grant No. JP21H01808. The authors thank Susan Coppersmith for fruitful discussions. Correspondence and requests for materials should be addressed to DK (dohunkim@snu.ac.kr).

**Author contributions**

DK conceived and supervised the project. YS and JK fabricated the device. YS, JY, and HJ performed the measurements and analyzed the data with WJ. JP, MC, and HS built the experimental setup and configured the measurement software. SM, NU, and KMI synthesized and provided the $^{28}$Si/SiGe heterostructure. All the authors contributed to the preparation of the manuscript.

**Competing interests**

The authors declare no competing interests.

# Figure captions

# Figure 1.

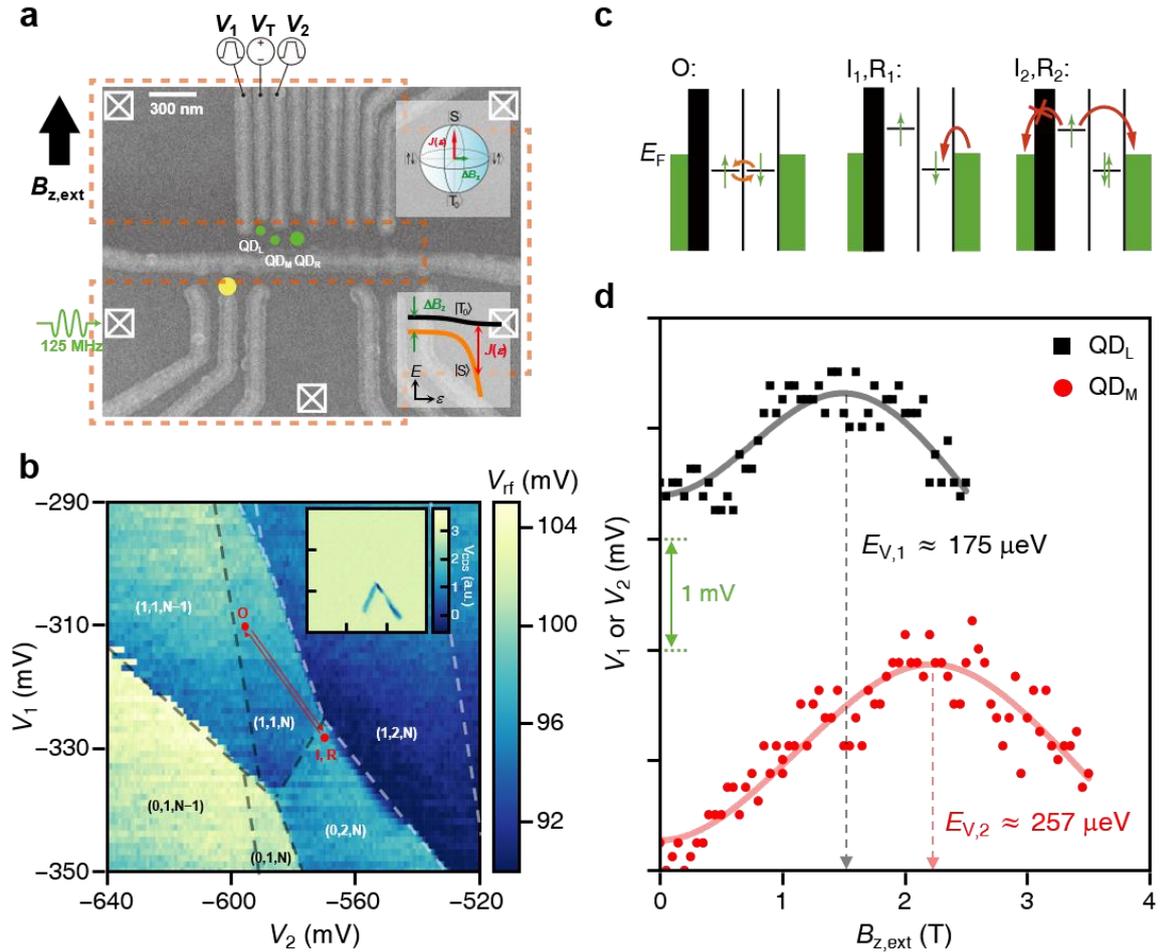

**Fig. 1 | The quantum dot device and triple quantum dot system.** (a) Scanning electron microscopy image of the device with the accumulation gates and Co micromagnet omitted. The black arrow indicates the direction of the external magnetic field $B_{z,\text{ext}}$. We focused on three quantum dots, indicated by the green dots labeled $QD_L$, $QD_M$, and $QD_R$. We used $QD_L$ and $QD_M$ as the $ST_0$ qubit and the many-electron dot $QD_R$ to explore the coherent interactions with the $ST_0$ qubit. High-frequency and synchronous voltage pulses combined with DC voltage were input to gates $V_1$, $V_2$ and $V_T$ to tune and manipulate the quantum systems. The yellow dot indicates the sensor dot based on an RF single-electron transistor, with a transpassing RF signal of ~ 125 MHz through RF Ohmic contact (indicated by the crossed squares). The orange dashed line indicates the micromagnet employed to apply a magnetic field difference $\Delta B_z$ between $QD_L$ and $QD_M$. The inset in the lower right corner illustrates the general energy level of the singlet and triplet states in a two-electron $ST_0$ qubit, with $\Delta B_z$ and detuning the ε-dependent exchange interaction $J(\varepsilon)$. The inset in the upper right corner depicts Bloch sphere representations of the contributions of $J(\varepsilon)$ and $\Delta B_z$ concerning the qubit rotation axis, with the two-electron states of the $ST_0$ qubit. (b) Charge stability diagram of the primary operational region for $QD_L$, $QD_M$, and $QD_R$. The number in parentheses represents the number of electrons in each of the three green dots. The inset shows $V_{CDS}$, the correlated double sampling signal of reflected RF signal $V_{rf}$. We drove the $ST_0$ qubit to reach I-O-R sequentially by applying appropriate pulse sequences with $V_1$ and $V_2$, while additional stopover points can be added to obtain the desired final qubit state. (c) Schematic of free

evolution of the $ST_0$ qubit in O and the initialize/readout sequence in the I/R points in (b), respectively. (d) Measurement of the valley splitting of $QD_L$ and $QD_M$ via magnetospectroscopy.

**Figure 2.**

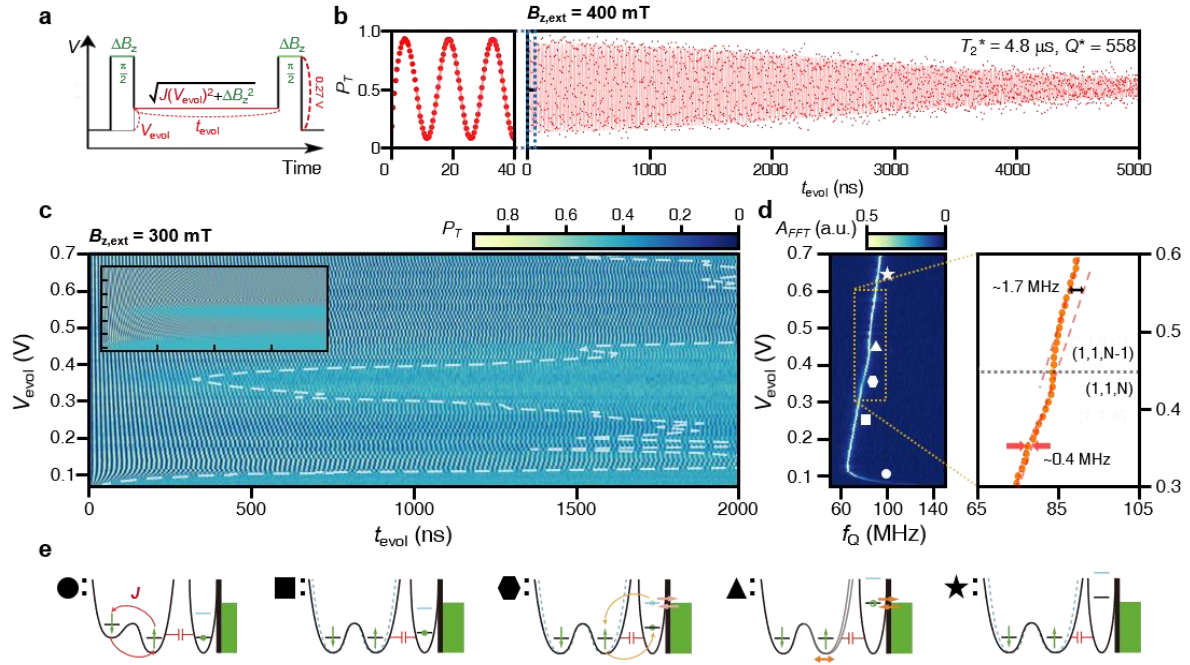

**Fig. 2 | The qubit dynamics revealed by Ramsey interferometry.** (a) Diagram of the pulse sequence used for Ramsey oscillation, the z-axis manipulation on the Bloch sphere with the free evolution time $t_{evol}$, and the pulse amplitude $V_{evol}$. A $\pi/2$ pulse was applied with $V_{evol}$ = 0.27 V and appropriately calibrated pulse duration time. (b) Representative Ramsey oscillation with the probability of the triplet state $P_T$ at $B_{z,ext}$ = 400 mT and $V_{evol}$ = 770 mV, with high coherence time $T_2^*$ and quality factor $Q^*$ values. The results on the left and right were averaged 10,000 and 100 times, respectively. (c) Ramsey oscillations as a function of $t_{evol}$ and $V_{evol}$. The white dashed line indicates the contour line of $T_2^*$ extracted from each Ramsey oscillation line of $V_{evol}$. The inset shows the numerically simulated Ramsey oscillation results. (d) Line-to-line FFT result of (c). The expected transition line of the dot on the right is indicated as a horizontal dotted line in the figure on the right. The two dashed orange lines show the linearly fitted $f_Q$ with $V_{evol}$ before and after the frequency shift of $\Delta f_Q \sim 1.7$ MHz during the charge transition of $QD_R$. A small bump with maximum $\Delta f_Q \sim 0.4$ MHz, indicated by the red arrows, is the footprint of the enhanced spin-dependent charge number fluctuation led by fast tunneling between $QD_R$ and the electron reservoir on the right side of $QD_R$. (e) Schematic depicting the energy levels of each marker in (d), capacitive coupling between the $ST_0$ qubit and $QD_R$, and the tunneling between $QD_R$ and the electron reservoir.

**Figure 3.**

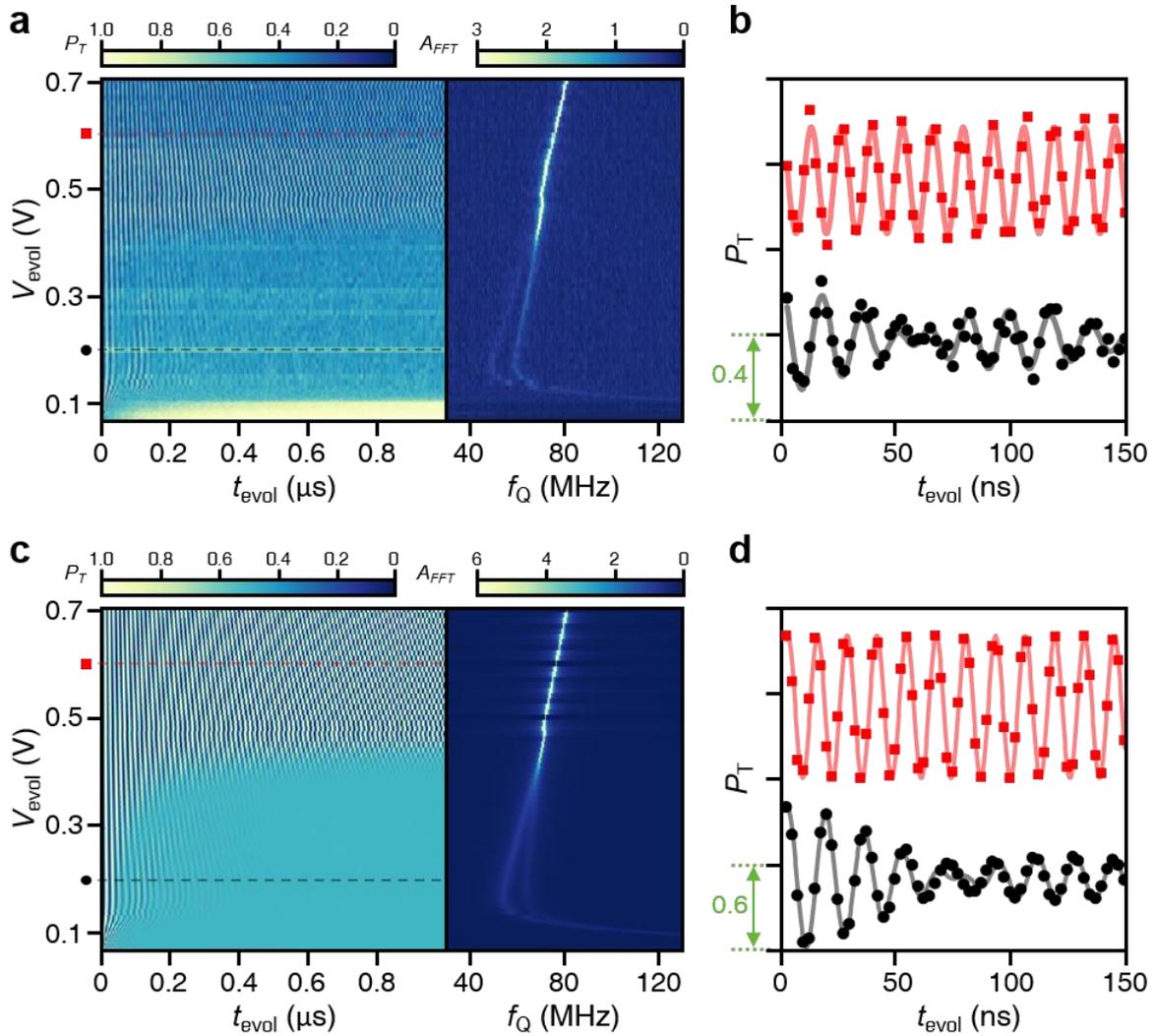

**Fig. 3 | Coherent coupling between the singlet-triplet qubit and many-electron spin states.** (a) Left: Ramsey oscillations as a function of $t_{evol}$ and $V_{evol}$ for different tuning levels. Significant dephasing appears below $V_{evol} = 0.42$ V. Right: FFT result of the figure on the left. The FFT peak exhibits the characteristic kink near $V_{evol} = 0.5$ V and linear dependence on $V_{evol}$ in the $\Delta B_z$-dominating regime. Characteristic splitting of the FFT peak is also manifested below $V_{evol} = 0.42$ V. (b) Ramsey oscillation trace at $V_{evol} = 0.6$ V (top, red squares) and 0.2 V (bottom, black solid circles). The beating of the oscillation at $V_{evol} = 0.2$ V is manifested. Each trace corresponds with the dashed line in the respective color in (a). The traces are offset by 1 for clarity. (c) Left: Numerical simulation of Ramsey oscillations as a function of $t_{evol}$ and $V_{evol}$. The significant decoherence below $V_{evol} = 0.42$ V was reproduced consistently. Right: FFT result of the figure on the left. The FFT peak shows the kink near $V_{evol} = 0.5$ V, a $V_{evol}$ dependence similar to the experimental results, and the characteristic splitting below $V_{evol} = 0.42$ V. (d) Simulated Ramsey oscillation trace at $V_{evol} = 0.6$ V (top, red squares) and 0.2 V (bottom, black solid circles). The simulated oscillation trace also reflects the beating of the oscillation at $V_{evol} = 0.2$ V. Each trace corresponds with the dashed line in the respective color in (c). The traces are offset by 1.2 for clarity.

**Figure 4.**

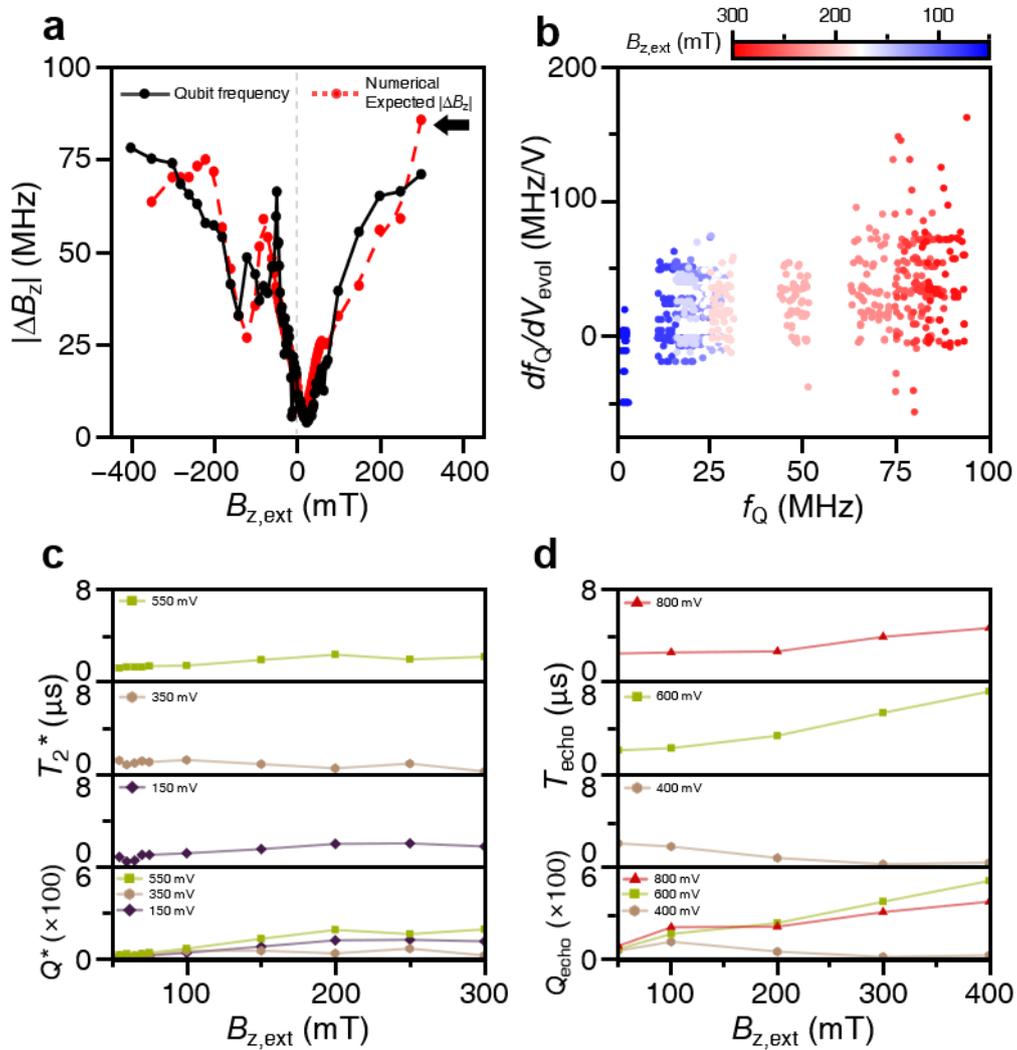

**Fig. 4 | External magnetic field dependence of coherence time.** (a) Estimated $|\Delta B_z|$ extracted from $f_Q$ of the Ramsey oscillation performed for $V_{evol}$ = 800 mV in the $\Delta B_z$-dominating region (black), and expected $|\Delta B_z|$ from the magnetic field simulation using OOMMF (red) (see Supplementary Information S5). The black arrow indicates the direction of the measurement. (b) Extracted $f_Q$ and charge susceptibility $df_Q/dV_{evol}$ of the qubit in $B_{z,ext}$ of 300–40 mT and $V_{evol}$ of 0.4–0.7 V. $df_Q/dV_{evol}$ was derived with interpolated $f_Q$. (c) Measured $T_2^*$, $Q^*$ of the Ramsey oscillations as a function of $B_{z,ext}$ at various $V_{evol}$. (d) $T_{echo}$ and $Q_{echo}$, the results of the spin-echo experiment of $T_2^*$ and $Q^*$, respectively, as a function of $B_{z,ext}$ at various $V_{evol}$.

**Figure 5.**

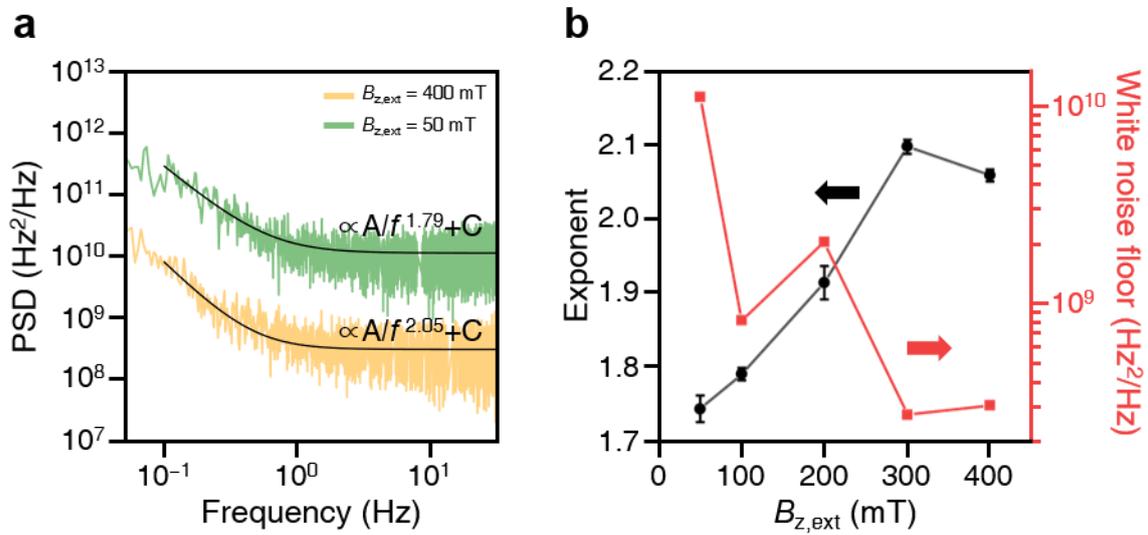

**Fig. 5 | Noise spectrum analysis.** (a) Noise spectrum acquired by applying two different $B_{z,ext}$. Power spectral densities (PSDs) were derived by analyzing single-shot data with the rapid Bayesian estimation method (see Supplementary Information S7). Each noise spectrum was calculated using 100,000 single shots. Offsets were excluded from this figure. (b) Power-law exponent and white noise floor level obtained from the noise spectrum at each $B_{z,ext}$.

**References**

1. Loss, D. & DiVincenzo, D. P. Quantum computation with quantum dots. *Phys. Rev. A* **57**, 120–126 (1998).

2. Petta, J. R. *et al.* Coherent Manipulation of Coupled Electron Spins in Semiconductor Quantum Dots. *Science* **309**, 2180–2184 (2005).

3. Koppens, F. H. L. *et al.* Driven coherent oscillations of a single electron spin in a quantum dot. *Nature* **442**, 766–771 (2006).

4. Nadj-Perge, S., Frolov, S. M., Bakkers, E. P. A. M. & Kouwenhoven, L. P. Spin–orbit qubit in a semiconductor nanowire. *Nature* **468**, 1084–1087 (2010).

5. Van Den Berg, J. W. G. *et al.* Fast Spin-Orbit Qubit in an Indium Antimonide Nanowire. *Phys. Rev. Lett.* **110**, 066806 (2013).

6. Pioro-Ladrière, M. *et al.* Electrically driven single-electron spin resonance in a slanting Zeeman field. *Nature Physics* **4**, 776–779 (2008).

7. Takeda, K. *et al.* A fault-tolerant addressable spin qubit in a natural silicon quantum dot. *Science advances* **2**, e1600694 (2016).

8. Scappucci, G. *et al.* The germanium quantum information route. *Nature Review Materials* **6**, 926–943 (2020).

9. Maurand, R. *et al.* A CMOS silicon spin qubit. *Nature Communications* **7**, 13575 (2016).

10. Jock, R. M. *et al.* A silicon metal-oxide-semiconductor electron spin-orbit qubit. *Nature Communications* **9**, 1768 (2018).

11. Maune, B. M. *et al.* Coherent singlet-triplet oscillations in a silicon-based double quantum dot. *Nature* **481**, 344–347 (2012).

12. Kawakami, E. *et al.* Electrical control of a long-lived spin qubit in a Si/SiGe quantum dot. *Nature Nanotechnology* **9**, 666–670 (2014).

13. Yoneda, J. *et al.* A quantum-dot spin qubit with coherence limited by charge noise and

fidelity higher than 99.9%. *Nature Nanotechnology* **13**, 102–106 (2018).

14. Wu, X. *et al.* Two-axis control of a singlet–triplet qubit with an integrated micromagnet. *Proceedings of the National Academy of Sciences* **111**, 11938–11942 (2014).

15. Takeda, K., Noiri, A., Yoneda, J., Nakajima, T. & Tarucha, S. Resonantly Driven Singlet-Triplet Spin Qubit in Silicon. *Phys. Rev. Lett.* **124**, 117701 (2020).

16. Jock, R. M. *et al.* A silicon singlet–triplet qubit driven by spin-valley coupling. *Nature Communications* **13**, 641 (2022).

17. Cai, X., Connors, E. J., Edge, L. F. & Nichol, J. M. Coherent spin–valley oscillations in silicon. *Nature Physics* **19**, 386–393 (2023).

18. Dumoulin Stuyck, N. I. *et al.* Low dephasing and robust micromagnet designs for silicon spin qubits. *Applied Physics Letters* **119**, (2021).

19. Reilly, D. J., Marcus, C. M., Hanson, M. P. & Gossard, A. C. Fast single-charge sensing with a rf quantum point contact. *Applied Physics Letters* **91**, 162101 (2007).

20. Noiri, A. *et al.* Radio-Frequency-Detected Fast Charge Sensing in Undoped Silicon Quantum Dots. *Nano Letters* **20**, 947–952 (2020).

21. Ono, K., Austing, D. G., Tokura, Y. & Tarucha, S. Current Rectification by Pauli Exclusion in a Weakly Coupled Double Quantum Dot System. *Science* **297**, 1313–1317 (2002).

22. Orona, L. A. *et al.* Readout of singlet-triplet qubits at large magnetic field gradients. *Phys. Rev. B* **98**, 125404 (2018).

23. Harvey-Collard, P. *et al.* High-Fidelity Single-Shot Readout for a Spin Qubit via an Enhanced Latching Mechanism. *Phys. Rev. X* **8**, 021046 (2018).

24. Nakajima, T. *et al.* Robust Single-Shot Spin Measurement with 99.5% Fidelity in a Quantum Dot Array. *Phys. Rev. Lett.* **119**, 017701 (2017).

25. Fogarty, M. A. *et al.* Integrated silicon qubit platform with single-spin addressability,

exchange control and single-shot singlet-triplet readout. *Nature Communications* **9**, 4370 (2018).

26. Delbecq, M. R. *et al.* Quantum Dephasing in a Gated GaAs Triple Quantum Dot due to Nonergodic Noise. *Phys. Rev. Lett.* **116**, 046802 (2016).

27. Connors, E. J., Nelson, J., Qiao, H., Edge, L. F. & Nichol, J. M. Low-frequency charge noise in Si/SiGe quantum dots. *Phys. Rev. B* **100**, (2019).

28. Struck, T. *et al.* Low-frequency spin qubit energy splitting noise in highly purified 28Si/SiGe. *npj Quantum Information* **6**, 40 (2020).

29. Shi, Z. *et al.* Tunable singlet-triplet splitting in a few-electron Si/SiGe quantum dot. *Applied Physics Letters* **99**, 233108 (2011).

30. Breuer, H.-P. & Petruccione, F. *The theory of open quantum systems*. (Clarendon, 2007).

31. Donahue, M. J. & Porter, D. G. OOMMF User's Guide, Version 1.0. (1999).

32. Neumann, R. & Schreiber, L. R. Simulation of micro-magnet stray-field dynamics for spin qubit manipulation. *Journal of Applied Physics* **117**, 193903 (2015).

33. Connors, E. J., Nelson, J., Edge, L. F. & Nichol, J. M. Charge-noise spectroscopy of Si/SiGe quantum dots via dynamically-decoupled exchange oscillations. *Nature Communications* **13**, 940 (2022).

34. Zhao, R. *et al.* Single-spin qubits in isotopically enriched silicon at low magnetic field. *Nature Communications* **10**, 5500 (2019).

35. Sergeevich, A., Chandran, A., Combes, J., Bartlett, S. D. & Wiseman, H. M. Characterization of a qubit Hamiltonian using adaptive measurements in a fixed basis. *Phys. Rev. A* **84**, 052315 (2011).

36. Shulman, M. D. *et al.* Suppressing qubit dephasing using real-time Hamiltonian estimation. *Nature Communications* **5**, 5156 (2014).

37. Nagaraja, H. S. *et al.* Magnetic Domain Studies of Cobalt Nanostructures. *J Supercond Nov Magn* **25**, 1901–1906 (2012).

38. Mi, X., Cady, J. V., Zajac, D. M., Deelman, P. W. & Petta, J. R. Strong coupling of a single electron in silicon to a microwave photon. *Science* **355**, 156–158 (2017).

39. Mi, X. *et al.* A coherent spin–photon interface in silicon. *Nature* **555**, 599–603 (2018).

40. Veldhorst, M. *et al.* An addressable quantum dot qubit with fault-tolerant control-fidelity. *Nature Nanotechnology* **9**, 981–985 (2014).

# Supplementary Information

## Supplementary Note 1. Signal-to-noise ratio (SNR) of the charge sensor

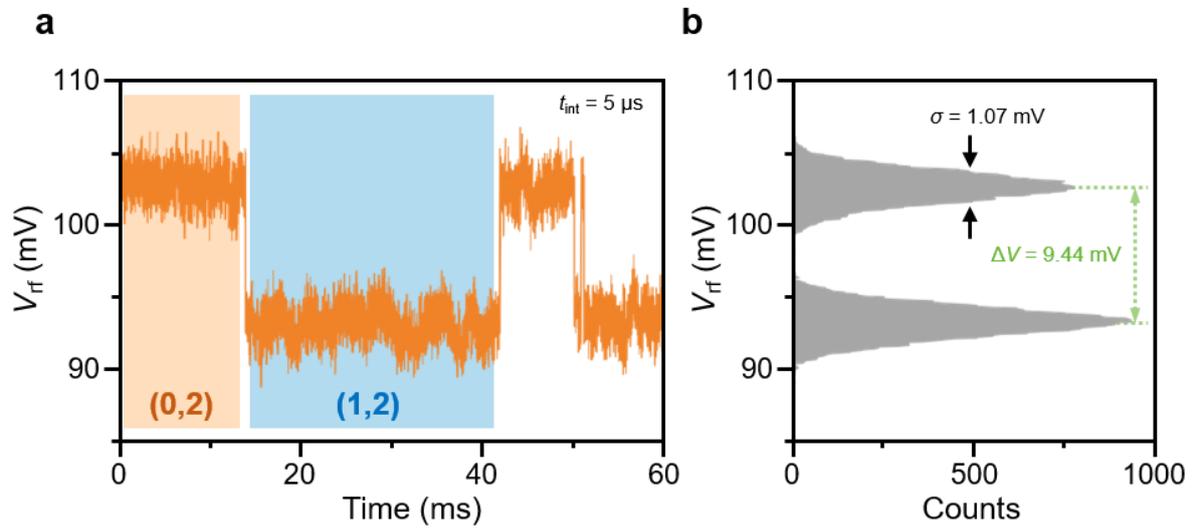

**Fig. S1 | Single electron charge transition signal.** (a) Demodulated sensor signal ($V_{rf}$) during single electron charge transitions from the left dot with integration time $t_{int}$ = 5 μs, and (b) its histogram. The SNR is defined by $\Delta V/\sigma$, where $\Delta V$ is the signal contrast of $V_{rf}$ for a single electron charge transition and $\sigma$ is the root-mean-square noise amplitude at a given $t_{int}$. $\Delta V$ and $\sigma$ are estimated to be 9.44 mV and 1.07 mV, respectively, yielding the SNR to be 8.82. This SNR alone would limit the state measurement infidelity < 1%, thus we regard the observed quantum oscillation visibility about 85% in the main text is limited by additional measurement error such as the triplet state relaxation during the readout and imperfect singlet state initialization after measurement.

# Supplementary Note 2. DC tuning dependency of the kink in the Ramsey interferometry

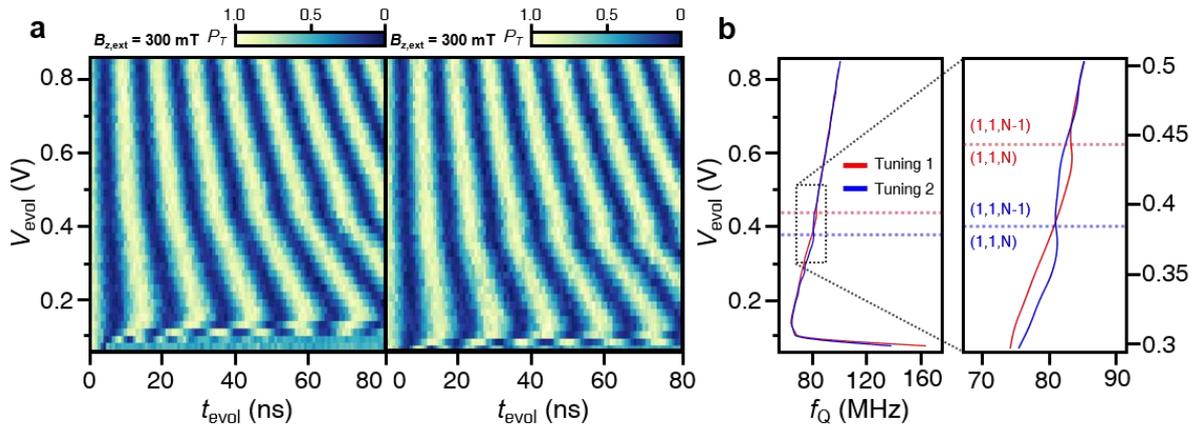

**Fig. S2 | Tunable kink in the Ramsey interferometry depending on the tuning condition.** (a) The Ramsey interferometry result under the DC voltage tuning condition same as in Fig. 2c of the main text (left panel, tuning 1) and with an additional –40 mV gate voltage applied to the plunger gate of $QD_R$ (right panel, tuning 2). The nearby gate voltages were appropriately tuned to enable state preparation and readout. (b) The interpolated line-to-line FFT result of (a), showing the systematic change of the kink point. The horizontal dotted lines indicate the kink point and the numbers in parenthesis represent the number of electrons of $QD_L$, $QD_M$, and $QD_R$, while both are shown with the red (blue) color for the data extracted under the tuning 1 (tuning 2) condition. The result indicates that the abrupt frequency shift in the Ramsey interferometry is correlated with the ground state charge transition in $QD_R$.

# Supplementary Note 3. Disappearance of the coherent coupling in weak magnetic fields

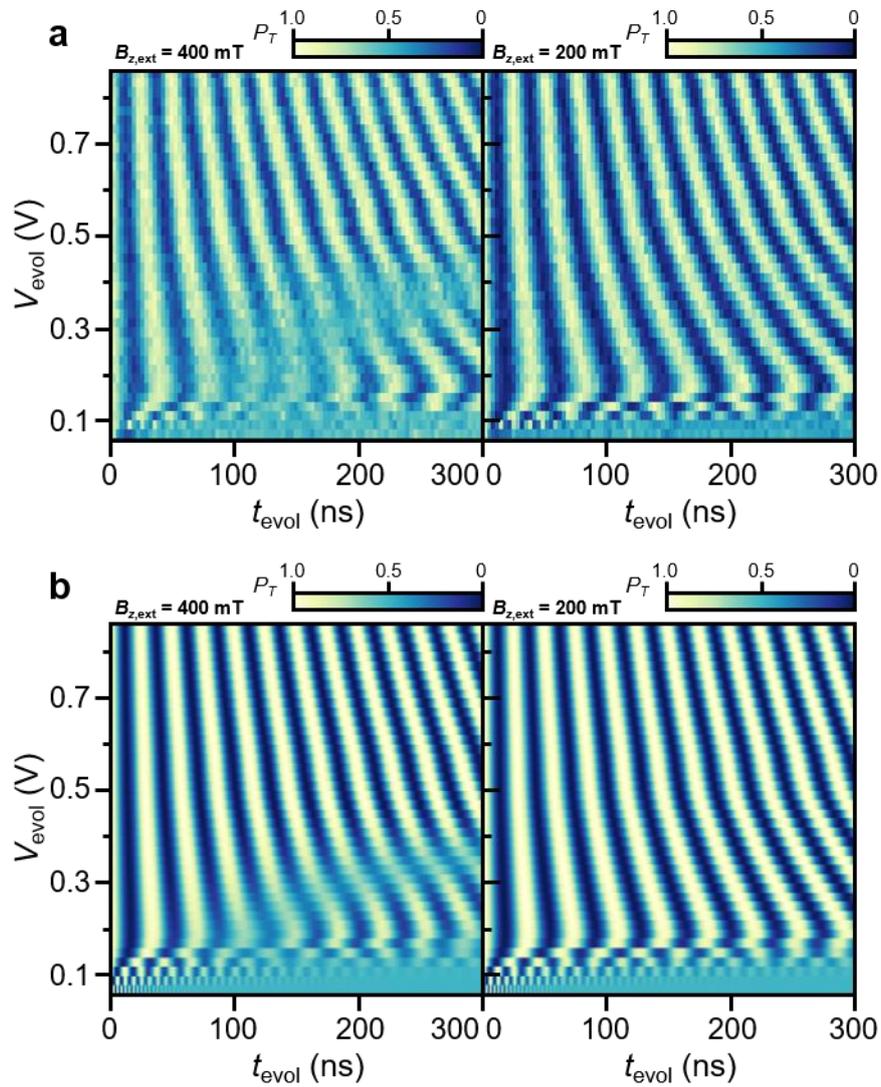

**Fig. S3 | Exchange spectroscopy in two different magnetic fields.** (a) The exchange spectroscopy result obtained using the pulse sequence depicted in Fig. 2a of the main text at two different magnetic fields (left: $B_{z,ext}$ = 400 mT, right: $B_{z,ext}$ = 200 mT). Note that the beating of the oscillation disappears as the magnetic field decreases, indicating that the eigenstate of the many-electron qubit depends on the magnetic field. (b) The simulation results obtained by solving the Lindblad master equation using the Hamiltonian described in Eqn. S1. The simulation results consistently reproduce the disappearance of the beating when the magnetic field is decreased from 400 mT to 200 mT, in support of the interpretation of the Zeeman-split spin levels in $QD_R$ as described in the main text.

**Supplementary Note 4. Details of the numerical simulation of the Ramsey interferometry**

In the main text, we implemented the Hamiltonian of the two interacting qubits to simulate the exchange spectroscopy result of the $ST_0$ qubit. Specifically, the Hamiltonian of the many-electron qubit ($QD_R$) that couples with $ST_0$ qubit is assumed to have a diagonal term with exponential detuning dependence and an off-diagonal term with linear magnetic field dependence. The coupling between the qubits is assumed to have the form of $J_{int}(V_{evol})\sigma_x \otimes \sigma_z$, and the coupling strength $J_{int}(V_{evol})$ is chosen to be the linear function of the detuning based on the spectroscopy result of Fig 3a of the main text. In addition, to incorporate the change in the charge state of the nearby N-electron quantum dot, the coupling term is multiplied by a phenomenological Fermi-Dirac distribution. Labeling $ST_0$ and $QD_R$ subspace with the subscript of S and R, the Hamiltonian in the Hilbert space $\mathcal{H}_S \otimes \mathcal{H}_R$ is expressed as below:

$$H(V_{evol}, B_{z,ext}) = J(V_{evol})\sigma_z \otimes \mathbf{1} + \Delta B_z(V_{evol}, B_{z,ext})\sigma_x \otimes \mathbf{1} + \beta(V_{evol})\mathbf{1} \otimes \sigma_z \\ + \gamma(B_{z,ext})\mathbf{1} \otimes \sigma_x + J_{int}(V_{evol})\frac{1}{e^{\eta(\beta(V_{evol})-c)}}\sigma_x \otimes \sigma_z \quad (S1)$$

, where $\mathbf{1}$ is a 2-dimensional identity matrix, $\beta(V_{evol}) = A_1 + A_2 \exp((A_3 - V_{evol})/\alpha)$ is the exponential function of detuning, and $\gamma(B_{z,ext}) = \kappa(B_{z,ext} - \delta)\theta(B_{z,ext} - \delta)$ is the linear function $\kappa$ of the magnetic field multiplied by the Heaviside step function $\theta$ for computational efficiency. Proper constants c, $\eta$, $A_{i=1,2,3}$, $\alpha$ and $\delta$ were used for enhancing the accuracy of numerical simulation.

Our specific choice for the form of the $QD_R$ Hamiltonian is primarily informed by the observations presented in Fig. S3a, where the beating of the oscillation of the $ST_0$ qubit disappears at certain $B_{z,ext}$ and $V_{evol}$ values. As these observations reveal the dependence on $B_{z,ext}$ and $V_{evol}$ of the eigenaxis of $QD_R$, we introduced the diagonal term $\beta(V_{evol})$ and the off-diagonal term $\gamma(B_{z,ext})$. To elucidate further, the off-diagonal term $\gamma(B_{z,ext})$ ensures a nonnegligible $\gamma(B_{z,ext})$ value when $B_{z,ext}$ reaches a sufficiently large magnitude. Concurrently, the diagonal term $\beta(V_{evol})$ imparts the exponential dependence on $V_{evol}$ to the eigenaxis of $QD_R$. Therefore, when $B_{z,ext}$ is large enough such that $\gamma(B_{z,ext})$ is nonnegligible, the eigenstates of the $QD_R$ align closely with the $\sigma_x$-eigenstates below a certain $V_{evol}$ threshold. In contrast, when $B_{z,ext}$ is small enough to the extent that $\gamma(B_{z,ext})$ is negligible, the eigenstates of the $QD_R$ remain the $\sigma_z$-eigenstates regardless of $V_{evol}$ values of interest. Since the $\sigma_x$-eigenstates are the superposition of the $\sigma_z$-eigenstates whose orbital wavefunction directly couples with the qubit frequency of $ST_0$ qubits, this gives rise to the beating of the oscillation at certain $V_{evol}$ and $B_{z,ext}$ values.

The simulation results of the exchange spectroscopy in the main text and supplementary information are obtained with the single qubit dephasing Lindbladian of $\tau_1\sqrt{J(V_{evol})}\sigma_z \otimes \mathbf{1}$ and $\tau_2\sqrt{\beta(V_{evol})}\mathbf{1}\otimes\sigma_z$ with dephasing rates $\gamma_S = J(V_{evol})\tau_1^2$ and $\gamma_M = \beta(V_{evol})\tau_2^2$. Here, assuming that the voltage fluctuation induces the dephasing of each qubit, we set $\gamma_S$ and $\gamma_M$ to be proportional to $J(V_{evol})$ and $\beta(V_{evol})$, respectively. In addition, throughout the simulation, the proportionality constants for $ST_0$ qubit and $J_{int}(V_{evol})$, which are accessible from the experimental results, were determined by fitting the formula to the experimental data. On the contrary, the inaccessible proportionality functions and constants for $QD_R$, such as $\kappa$ and $\alpha$, were selected to give the results that fit the experimental data well. The exchange spectroscopy was simulated by numerically solving the Quantum Master equation (QME) in a Lindblad form, with $V_{evol}(t)$ of the pulse sequence depicted in Fig. 2a of the main text.

## Supplementary Note 5. Numerical micromagnetic simulation

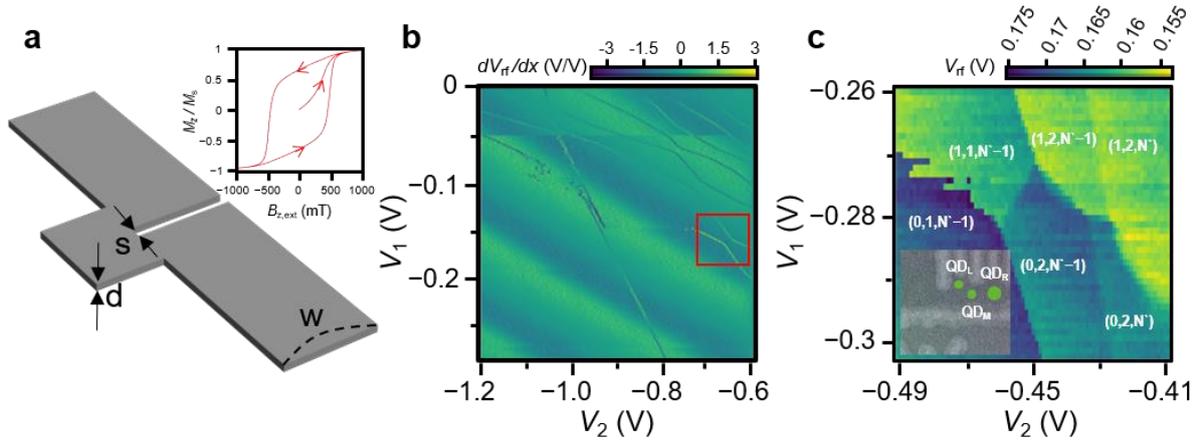

**Fig. S4 | On-chip Co Micromagnet.** (a) The design of the micromagnet used in our study. For simulation, we set the parameters with thickness d = 128 nm, width w = 2.112 μm and gap s = 288 nm. Inset: The hysteresis curve with magnetization in z-direction $M_z/M_s$, with $B_{z,ext}$ swept between +2000 mT and -2000 mT starting at 0 mT. (b) Charge stability diagram while the plunger gate and right barrier gate of $QD_R$ are fully merged with the right reservoir, indicating that $QD_L$ and $QD_M$ for the $ST_0$ qubit reside inside the region formed by the three left most surface gates. The red box indicates the primary operational region. (c) Charge stability diagram after forming the third, many-electron quantum dot $QD_R$.

We used Object Oriented Micromagnetic Framework (OOMMF) to calculate numerically expected $|\Delta B_z|$, which solves Landau-Lifshift-Gilbert equation[1] for each time step to obtain the magnetization in each mesh with a given magnetic field. We modeled the 3-dimensional shape of the micromagnet (Fig. S4a) we used in this study, consisting of polycrystalline cobalt[2]. The magnetization process was done by driving $B_{z,ext}$ parameter back and forth from 500 mT to –500 mT several times, similar to the magnetization process we adopted. The following parameters were used in the simulations: Uniaxial anisotropy constant $K_1$ = 7.5×10⁵ J/m³, $K_2$ = 1.5×10⁵ J/m, exchange constant A = 2.906×10¹¹ J/m, saturation magnetization $M_s$ = 1.467×10⁶ A/m, Landau-Lifshitz gyromagnetic ratio $\gamma_{LL}$ = 2.409×10⁵ m/(As), Gilbert damping parameter 0.5, and the cubic mesh size = 32 nm. Some of the parameters were referred from the previous study[3] which analyzed this micromagnet design with numerical simulation earlier, and we could reproduced similar hysteresis loop (see inset of Fig. S4a) with aforementioned study using those parameters.

For calculating numerically expected $|\Delta B_z|$, we used Gaussian averaged $B_z$, with diameter at 25nm and its center at the expected position of $QD_L$ and $QD_M$. To explore the expected position of $QD_L$ and $QD_M$, we formed the quantum dot using only the three left most gates and backbone gates (see Fig. S4b). The charge stability diagram indicates that $QD_L$ and $QD_M$ are likely located inside the region formed by the three left most gates. Subsequent formation of $QD_R$ and its coupling with the $ST_0$ qubit is shown in Fig. S5c. We selected the expected position of each of two quantum dots (see inset of Fig. S4c) where the numerically expected $|\Delta B_z|$ results most fit the measured $|\Delta B_z|$ result, and this dot position also showed consistency from the actual experiment, which we revealed from Fig. S4b and Fig. S4c. The numerically expected distance between $QD_L$ and $QD_M$ was ~ 100nm. The vertical distance between the bottom of the magnet and the quantum dots for simulation was 163nm, estimated from the device structure.

# Supplementary Note 6. Effect of total data acquisition time on the coherence time

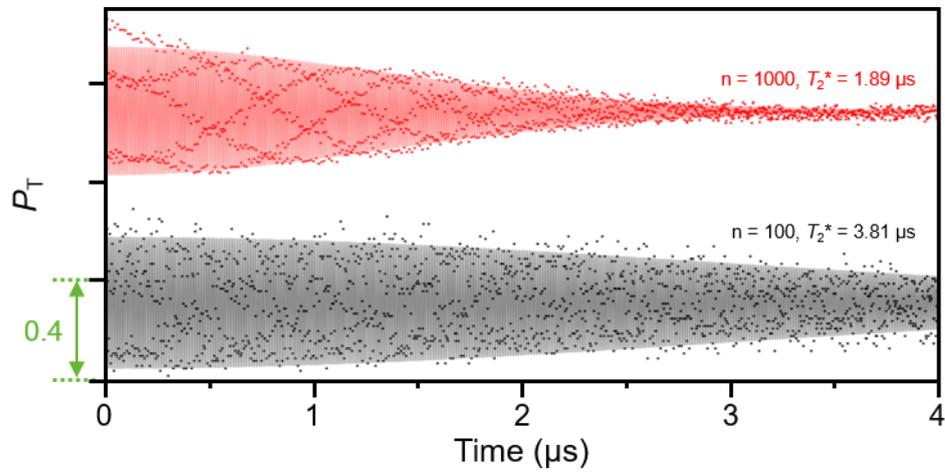

**Fig. S5 | Ramsey oscillation with different data acquisition times.** The magnetic field driven qubit oscillation traces of $ST_0$ qubit at $B_{z,\text{ext}}$ = 400 mT with the number of single-shot experiments for each data point n = 100 (lower trace) and n = 1000 (upper trace). The traces are offset by 0.7 for clarity. Even at relatively strong $B_{z,\text{ext}}$ = 400 mT, increasing total data acquisition time leads to reduced the coherence time indicating the significant portion of low frequency noise in the system.

**Supplementary Note 7. Bayesian frequency estimation and power spectral density**

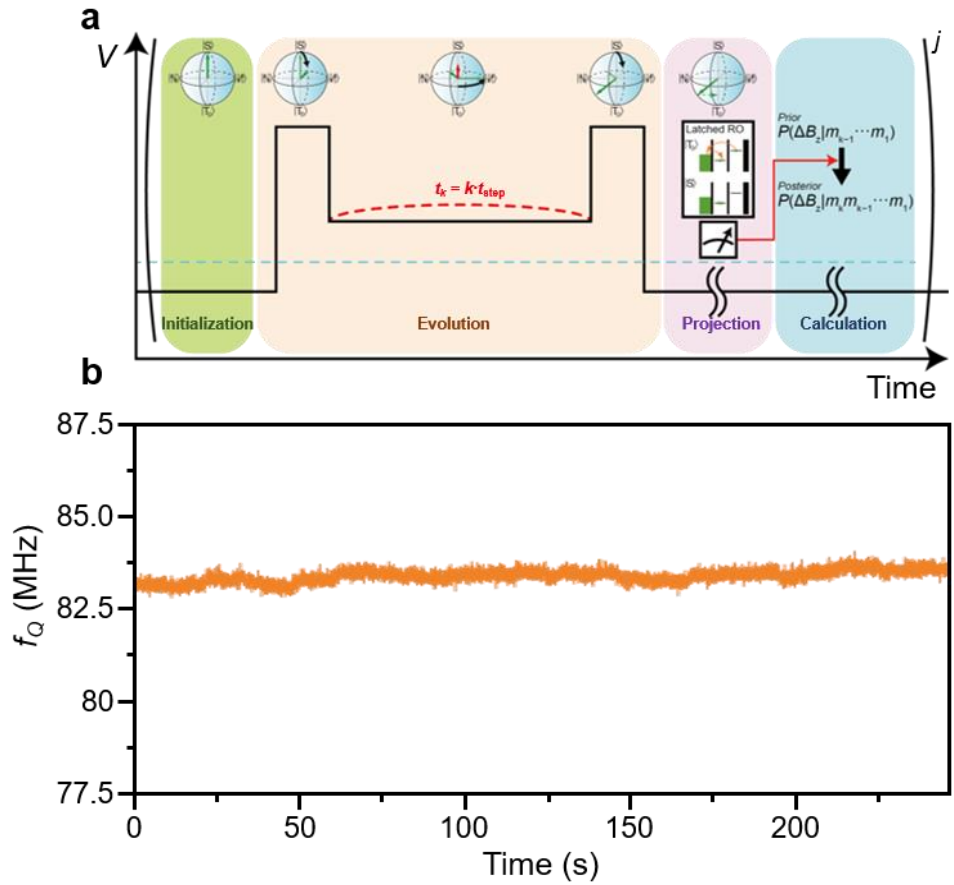

**Fig. S6 | Bayesian estimation for ST$_0$ qubit.** (a) Schematic of Bayesian estimation circuit for ST$_0$ qubit, used in our study. (b) Example of the time trace of $f_Q$ using the circuit (a) at $B_{z,\text{ext}}$ = 400 mT.

The Bayesian estimation circuit, illustrated in Fig. S6a, starts with the initialization step, where we first initialize ST$_0$ qubit to the singlet state. This step is followed by the evolution step, where we manipulate the qubit at a specific detuning for free evolution $V_{\text{evol}}$ with a given evolution time $t_k$ that linearly increase in step of $t_{\text{step}}$ up to a given number of maximum iteration $j$. The subsequent projective single-shot measurement of ST$_0$ qubit is performed by the latched readout method described in the main text. Based on the single-shot measurement result, Bayesian inference of the qubit frequency $f_Q$ is performed by the following rule up to a normalization constant:

$$P(f_Q | m_j, m_{j-1}, ..., m_1) = P_0(f_Q) \prod_{k=1}^{j} \frac{1}{2}\{1 + r_k[\alpha + \beta \cos(2\pi f_Q t_k)]\}$$

, where $r_k = 1(-1)$ for $m_k = |S\rangle(|T_0\rangle)$, and $\alpha(\beta)$ is the parameter determined by the axis of rotation (oscillation visibility). Starting from the prior uniform initial distribution $P_0(f_Q)$, the narrow posterior distribution $P(f_Q | m_j, m_{j-1}, ..., m_1)$ is obtained after $j^{th}$ Bayesian estimation and $f_Q$ is estimated where the probability value is the maximum. In our study, we set $\alpha = -0.05$, $\beta = 0.6$, $t_{step} = 1.6$ ns and $j = 200$ for Bayesian estimation. The time trace of $f_Q$ (see Fig. S6b, as an example) is converted to power spectral density in the frequency domain by fast Fourier transform and the Welch's method[4] was used for smoothing the spectrum.

**Supplementary References**

1. Gilbert, T. L. Classics in Magnetics A Phenomenological Theory of Damping in Ferromagnetic Materials. *IEEE Trans. Magn.* **40**, 3443–3449 (2004).

2. Nagaraja, H. S. *et al.* Magnetic Domain Studies of Cobalt Nanostructures. *J Supercond Nov Magn* **25**, 1901–1906 (2012).

3. Neumann, R. & Schreiber, L. R. Simulation of micro-magnet stray-field dynamics for spin qubit manipulation. *Journal of Applied Physics* **117**, 193903 (2015).

4. Welch, P. The use of fast Fourier transform for the estimation of power spectra: A method based on time averaging over short, modified periodograms. *IEEE Trans. Audio Electroacoust.* **15**, 70–73 (1967).